\documentstyle[12pt,epsf]{article}

\begin{document}
\sloppy
\sloppy
\sloppy
$\ $
\begin{flushright}{UT-759,1996}
\end{flushright}
\vskip 0.5 truecm

\vskip 0.5 truecm

\begin{center}
{\large{\bf  A toy model of quantum gravity with\\ a quantized cosmological
constant\footnote{To be published in Progress of Theoretical Physics(Kyoto)} }}
\end{center}
\vskip .5 truecm
\centerline{\bf Kazuo Fujikawa}
\vskip .4 truecm
\centerline {\it Department of Physics,University of Tokyo}
\centerline {\it Bunkyo-ku,Tokyo 113,Japan}
\vskip 0.5 truecm

\vskip 0.5 truecm

\begin{abstract}
The path integral for the hydrogen atom, for example,  is formulated as a one-dimensional 
quantum gravity coupled to matter fields representing the electron coordinates.
The (renormalized) cosmological constant, which corresponds to the energy eigenvalue, is 
thus quantized in this model. Possible implications of the quantized 
cosmological constant are briefly  discussed.
\par

\end{abstract}

The problem of the cosmological constant in quantum gravity has a long history. For a review of this subject, see ref.[1] and papers cited therein.
Recently, we pointed out that the path integral of the hydrogen atom, for 
example,  in non-relativistic quantum mechanics is formulated as a 
one-dimensional quantum
gravity coupled to matter fields representing the electron coordinates.
The basic idea of this formulation is to use the Jacobi's principle of least action for the evaluation of the Green's function with a given energy[2]. In this formulation, the 
(renormalized) cosmological constant corresponds to the energy eigenvalue 
of the Schroedinger equation, and thus the cosmological constant is quantized.

Since the notion of a quantized cosmological constant appears to be something novel, at
least to my knowledge, we would like to present a brief description of the 
model. The model as it stands is a one-dimensional one and thus trivial 
as quantum gravity, but this model might have some interesting implications in 
the context of mini-superspace models, for example.

The model is defined by    
\begin{equation}
\int \frac{{\cal D}\vec{x}{\cal D}h}{gauge\  volume} exp\{ i\int_{0}^{\tau} Lhd\tau\}
\end{equation}
with 
\begin{equation}
L = \frac{m}{2h^{2}}(\frac{d\vec{x}}{d\tau})^{2} - V(r) + E
\end{equation}
where $h$ stands for the einbein, a one-dimensional analogue of vierbein $h_{\mu}^{a}$, and $h= \sqrt{g}$ in one-dimension. The variable $h$ stands for the lapse function $N$ in the conventional notation of quantum gravity.
We set the Planck constant $\hbar = 1$. The  matter 
variables $\vec{x}$ stand for the electron coordinates in the context of the 
hydrogen atom.
 This problem is analogous to the Polyakov-type path integral in string theory[3] - [5]. It is shown that the above path integral is re-written as the one 
with  canonical Liouville measure , up to a renormalization of the cosmological constant ( or  energy eigenvalue E).

The BRST invariant path integral for  (1)  is written by using the Faddeev-Popov procedure as 
\begin{eqnarray}
&&\int {\cal D}(\sqrt{h}\vec{x}){\cal D}\sqrt{h}{\cal D}(h^{3/2}c){\cal D}\bar{c}{\cal D}B \nonumber\\
&&\times exp\{i \int_{0}^{\tau}  L h d\tau + i\int_{0}^{\tau} [ B (\sqrt{h} - \sqrt{f}) - i \frac{1}{2}\bar{c}\sqrt{h}\partial_{\tau}c 
         - i \bar{c}c\partial_{\tau}(\sqrt{h} - \sqrt{f})] d\tau\}
\nonumber\\
&& \equiv \int d\mu \exp [ i\int_{0}^{\tau} {\cal L}_{eff} d\tau]
\end{eqnarray}
where the gauge condition is specified by $h(\tau) = f(\vec{x}(\tau))$.
The BRST transformation is defined as a translation in the Grassmann parameter
, $\theta \rightarrow \theta + \lambda$, in the superfield notation(note that 
$\theta^{2} = \lambda^{2} = \theta\lambda + \lambda\theta = \theta c(\tau) +
c(\tau)\theta = 0$)
\begin{eqnarray}
\vec{x}(\tau, \theta) &=& \vec{x}(\tau) + i \theta c(\tau)\partial_{\tau}\vec{x}(\tau)\nonumber\\
\sqrt{h(\tau, \theta)} &=& \sqrt{h(\tau)} + 
i\theta [c(\tau)\partial_{\tau} + \frac{1}{2}(\partial_{\tau}c(\tau))]\sqrt{h(\tau)}\nonumber\\
\sqrt{h}\vec{x}(\tau, \theta) &=& \sqrt{h}\vec{x}(\tau) + i \theta [ c(\tau)\partial_{\tau} + \frac{1}{2}(\partial_{\tau}c(\tau))]\sqrt{h}\vec{x}(\tau)\nonumber\\
\sqrt{f(\vec{x}(\tau, \theta))} &=& \sqrt{f(\vec{x}(\tau))} + i \theta c(\tau)
\partial_{\tau} \sqrt{f(\vec{x}(\tau))}\nonumber\\
c(\tau, \theta) &=& c(\tau) + i\theta c(\tau)\partial_{\tau}c(\tau)\nonumber\\
\bar{c}(\tau, \theta) &=& \bar{c}(\tau) + \theta B(\tau)
\end{eqnarray}
For example, the BRST transformation is given by 
\begin{eqnarray}
\delta \sqrt{h(\tau)} &=& i\lambda [c(\tau)\partial_{\tau} + \frac{1}{2}(\partial_{\tau}c(\tau))]\sqrt{h(\tau)}\nonumber\\
\delta (\sqrt{h}\vec{x}(\tau)) &=& i\lambda [c(\tau)\partial_{\tau} + \frac{1}{2}(\partial_{\tau}c(\tau))]\sqrt{h}\vec{x}(\tau)\nonumber\\
\delta (h^{3/2}dc(\tau)) &=& i\lambda [c(\tau)\partial_{\tau} + \frac{1}{2}(\partial_{\tau}c(\tau))](h^{3/2}dc(\tau))
\end{eqnarray}
In the last relation for the ghost variable $c(\tau)$, we consider the differential $dc(\tau)$:  one may write $h^{3/2}dc =  d(h^{3/2}c)$ for a fixed metric $h(\tau)$, which is the case 
required to study the BRST invariance of the path integral measure in (3).   Note that all the variables in (5) have the same BRST transformation law which is anomaly free[6]; the combination ${\cal D}\bar{c}{\cal D}B$ is also 
manifestly BRST invariant. These properties  in turn ensure the BRST
invariance of the path integral measure in (3). One can also confirm that the
action in (3) is BRST invariant.

The path integral (3) is rewritten as 
\begin{equation}
\int {\cal D}\tilde{\vec{p}}{\cal D}\tilde{\vec{x}}{\cal D}\sqrt{h}
{\cal D}\tilde{c}{\cal D}\bar{c}{\cal D}B 
exp\{ i\int_{0}^{\tau} d\tau [( \vec{p}\dot{\vec{x}} - Hh) +
B(\sqrt{h} - \sqrt{f}) - i\frac{1}{2}\bar{c}\sqrt{h}\partial_{\tau}
( \frac{1}{h^{3/2}}\tilde{c})]\}
\end{equation} 
with 
\begin{eqnarray}
H &\equiv& \frac{1}{2m}\vec{p}^{2} + V(r) - E \nonumber\\
\tilde{\vec{p}} &=& \sqrt{h}\vec{p}\nonumber\\
\tilde{\vec{x}} &=& \sqrt{h}\vec{x}\nonumber\\
\tilde{c} &=& h^{3/2}c
\end{eqnarray}
In fact, after the path integral over $\tilde{\vec{p}}$, one recovers (3). 
We also set $\sqrt{h} - \sqrt{f} = 0$ after a partial integration in the ghost 
sector in (6). Note that the path integral
\begin{equation}
\int {\cal D}\tilde{\vec{p}}\exp\{  i\int [- \frac{h}{2m}{\vec{p}}^{2} 
+ \vec{p}\dot{\vec{x}} - \frac{m}{2h}{\dot{\vec{x}}}^{2}] d\tau\}
= \int {\cal D}\tilde{\vec{p}}\exp [ i\int \frac{-1}{2m}(\tilde{\vec{p}})^{2} d\tau]
\end{equation}
is a constant independent of the metric $h(\tau)$: This is another way to see why the variables with weight $1/2$ such as $\tilde{\vec{p}}$ and $\tilde{\vec{x}}$ are chosen for the world 
scalar quantities $\vec{p}$ and $\vec{x}$  as reparametrization
invariant path integral variables.

In the above path integral, it is important to recognize that the singular (time-derivative) terms in the Lagrangian (6) are written as 
\begin{equation}
\int {\cal D}\tilde{\vec{p}}{\cal D}\tilde{\vec{x}}
{\cal D}\tilde{c}{\cal D}\bar{c} \exp\{  \int_{0}^{\tau} d\tau [ i\tilde{\vec{p}}\frac{1}{\sqrt{h}}\partial_{\tau}( \frac{1}{\sqrt{h}}\tilde{\vec{x}}) + \frac{1}{2}\bar{c}\sqrt{h}\partial_{\tau}
( \frac{1}{h^{3/2}}\tilde{c})]\}
\end{equation}
Those singular terms have a Weyl invariant structure; namely, the $h$-dependence 
can be completely removed by a suitable scale transformation of $\tilde{\vec{p}},\tilde{\vec{x}},\tilde{c}$ and $\bar{c}$ such as $ \tilde{\vec{x}}\rightarrow \sqrt{h}\tilde{\vec{x}}$ and $\bar{c}\rightarrow (1/\sqrt{h})\bar{c}$. In this process of scaling, one obtains a Jacobian (or anomaly), which can be integrated to a Wess-Zumino term[7]. At the same time, the path integral variables $\tilde{\vec{p}}$ and $\tilde{\vec{x}}$ are reduced to the naive ones. In the present case, one can confirm that the Jacobian (Weyl anomaly) has a form
\begin{equation}
M \int  h\delta\beta (\tau)d\tau
\end{equation}
for an infinitesimal scale transformation
\begin{eqnarray}
\bar{c}(\tau)&\rightarrow&  e^{- \delta\beta (\tau)}\bar{c}(\tau)\nonumber\\
\tilde{c}(\tau)&\rightarrow&  e^{3 \delta\beta (\tau)}\tilde{c}(\tau)\nonumber\\
\tilde{\vec{p}}(\tau)&\rightarrow&  e^{\delta\beta (\tau)}\tilde{\vec{p}}(\tau)\nonumber\\
\tilde{\vec{x}}(\tau)&\rightarrow&  e^{\delta\beta (\tau)}\tilde{\vec{x}}(\tau)
\end{eqnarray}
This evaluation of the Jacobian is performed for $\tilde{\vec{p}}$ and $\tilde{\vec{x}}$, for example, by[4][8]
\begin{eqnarray}
&&\lim_{M \rightarrow \infty}\int\frac{dk}{2\pi}e^{-ik\tau}\exp\{(\frac{1}{\sqrt{h}}\partial_{\tau}\frac{1}{\sqrt{h}})^{\dagger}(\frac{1}{\sqrt{h}}\partial_{\tau}\frac{1}{\sqrt{h}})/M^{2}\}e^{ik\tau}\nonumber\\
&&= \lim_{M \rightarrow \infty}M\int\frac{dk}{2\pi}\exp\{[\frac{1}{\sqrt{h}}(\partial_{\tau}/M + ik)\frac{1}{\sqrt{h}}][\frac{1}{\sqrt{h}}(\partial_{\tau}/M + ik)\frac{1}{\sqrt{h}}]\}\nonumber\\
&& = \lim_{M \rightarrow \infty}M\int\frac{dk}{2\pi}\exp\{ -k^{2}/h^{2}\}( 1 + 
O(\frac{1}{M^{2}}))\nonumber\\ 
&& = \frac{1}{2\sqrt{\pi}}M h \ \ \ \ for \ \ M\rightarrow \infty
\end{eqnarray}
which gives a term of a general structure as in (10).
The anomaly calculation is specified by the basic operators appearing in (9) 
\begin{equation}
\frac{1}{\sqrt{h}}\partial_{\tau}\frac{1}{\sqrt{h}}\ \ \ , or \ \ 
\sqrt{h}\partial_{\tau}\frac{1}{h^{3/2}}
\end{equation}
both of which give the same form of anomaly proportional to $h$,  as in (12).
See also Ref.[4]. The overall sign of the Jacobian is specified depending on the statistics of 
each variable, i.e., a Grassmann variable gives an extra minus sign. Since only the most singular term survives in (12), one can confirm that the knowledge of the most singular terms in the Lagrangian (6), i.e.,the time-derivative terms in  (9), is sufficient to 
calculate anomaly. If one denotes the integrated anomaly ( Wess-Zumino term) by $\Gamma (h)$, we  have from (10) 
\begin{equation}
\Gamma (h) - \Gamma (he^{-2\delta\beta}) = \int 2\delta\beta (\tau)h(\tau)\frac{\partial\Gamma}{\partial h(\tau)}d\tau = M\int h(\tau)\delta\beta (\tau)d\tau
\end{equation}
since $h$ is transformed to $he^{-2\delta\beta}$ by (11), and  we obtain
\begin{equation}
\Gamma (h) = M\int h(\tau)d\tau
\end{equation}
with  a suitable (infinite) number $M$. This $\Gamma (h)$  is added to the action in (6) and it has a form of the cosmological constant (or energy) in (6).

This calculation of the anomaly is quite general: In two-dimensions, the Weyl anomaly has a structure $M^{2}\sqrt{g} + \sqrt{g}R$,[3] - [5]. In one-dimension, the curvature term containing $R$ does not exist, and only the cosmological term arises if one performs a reparametrization invariant calculation. This Weyl anomaly  renormalizes the bare cosmological term in (6),
\begin{equation}
E + M = E_{r}
\end{equation}
with $E_{r}$ a renormalized cosmological constant.
But this renormalization is independent of the choice of the gauge fixing function $f(\vec{x}(\tau))$ in (3).

After this procedure of eliminating the $h$-dependence from  the time-derivative 
terms  and then the integration over $B$ and $h$, one obtains  a  path integral
\begin{equation}
\int {\cal D}\vec{p}{\cal D}\vec{x}
{\cal D}c{\cal D}\bar{c} 
exp\{ i\int_{0}^{\tau} d\tau [ \vec{p}\dot{\vec{x}} - f(\vec{x}(\tau))H  -  i\frac{1}{2}\bar{c}\partial_{\tau}c]\} 
\end{equation}
The {\em decoupled} ghost sector, which is first order in $\partial_{\tau}$, 
does not contribute to the Hamiltonian.

A convenient way to find the quantized cosmological constant is to consider the  Green's function 
\begin{equation}
\langle \vec{x}_{b}|\frac{1}{{\hat{\vec{p}}}^{2} + \hat{V}(r)  - E_{r} }|\vec{x}_{a}\rangle =
i\int_{0}^{\infty} d\tau_{f} \int {\cal D}\vec{p}{\cal D}\vec{x} 
exp\{ i\int_{0}^{\tau_{f}} d\tau [ \vec{p}\dot{\vec{x}} - f(\vec{x}(\tau))H ]\}
\end{equation}
by adjusting an over-all normalization factor in (17), which is independent of $f(\vec{x})$ but can depend on $\tau_{f}$. The Green's function is shown to be gauge 
independent (i.e., independent of the choice of $f(\vec{x})$, if the operator 
ordering is properly taken into account)[2][9], as is 
indicated in (18).

For the hydrogen atom, one may choose a gauge   
\begin{equation}
f(\vec{x}(\tau)) = r = \xi + \eta
\end{equation}
in the parabolic coordinates, and one can perform an exact path integral[9][10][2]   to recover the quantized (renormalized)  cosmological constant $E_{n} = - (1/2)m\alpha^{2}/n^{2}$, $n = 1, 2, ..$, for negative $E_{r}$ as poles in the Green's function. For positive $E_{r}$, the 
cosmological constant can be continuous. For other models of $V(r)$ such as
3-dimensional harmonic oscillator, one can derive an exact result by choosing
$f(\vec{x}) = 1$. Each eigenstate of the cosmological constant corresponds to an 
allowed physical  ``universe''. The quantization of the cosmological constant( energy
eigenvalue)  is thus quite common in the present context. An analogue of 
the Wheeler-DeWitt equation in the present problem is
\begin{equation}
\hat{H}_{T}\Psi (\vec{x}) =0
\end{equation}
with a total Hamiltonian $\hat{H}_{T} = \hat{f}(\vec{x})\hat{H}$ in the sense 
of Dirac. See ref.[2].  A suitable boundary condition on $\Psi (\vec{x})$ leadts to the quantization of a cosmological constant. If one requires the spherical
symmetry of $\Psi (\vec{x})$,for example,  the eigenvalues corresponding to $S$-states are
selected.

We here comment on an interesting aspect of the present problem from a view point of BRST symmetry. The partition function $Tr e^{-i\hat{H}\tau}$ corresponding to (3), which is defined by imposing periodic boundary conditions on all the 
variables including ghosts, is formally shown to be gauge independent for any 
value of the cosmological constant $E$. However, in reality the partition function is not quite gauge independent for a general $E$. This is related to the fact that the BRST invariant physical state ( or physical universe) is specified 
by a quantized cosmological constant and that the quantized cosmological constant selects only a specific set of states, for example $n^{2}$ degenerate states for a fixed $E_{n}$ in the Coulomb potential, as allowed physical states. This  property may have some relevance if a universe with a quantized cosmological
constant should be  realized.

It is hoped that the present mechanism of quantizing the cosmological 
constant may have some physical implications in a 
simplified model of quantum gravity such as mini-superspace models. More
generally, a suitable choice of the boundary condition on the Wheeler-DeWitt 
equation [11] might lead to a quantized (renormalized) cosmological constant,
which, if realized, will influence our way of thinking about the cosmological 
constant. The importance of the boundary condition on the Wheeler-DeWitt
equation or related path integral  has been emphasized by Weinberg; without boundary conditions being
specified, quantum cosmology is an incomplete theory [1]. The proposal of Hawking [12] and
Coleman [13], according to which the wave function of the universe has a 
sharp peak at a very specific value of the cosmological constant, might be 
regarded as describing a different aspect of the quantization of the 
cosmological constant: At least, the fact that the wave function of the ``universe'' has a $\delta$ functional peak for a very specific value of the cosmological constant is common to both of the Hawking-Coleman mechanism and the present quantized cosmological constant.\\
\\
{\bf Note added}\\
After submitting the present paper for publication, the papers in ref.(14), where some related technical matters are treated, came to my attention. However, the notion of a quantized cosmological constant is not discussed 
in these papers. I thank Dr. M. Seriu for calling these papers to my attention.

\end{document}